\documentclass[a4paper]{article}
\usepackage[utf8]{inputenc}
\usepackage[english]{babel}
\usepackage{amsmath}
\usepackage{authblk}
\usepackage{graphicx}
\usepackage{units}
\usepackage[top=2.5cm,left=2.5cm,right=2.5cm,bottom=2.5cm]{geometry}
\usepackage{url}

\title{Prospects of Event Shape Sorting}

\author[1]{Boris Tom\' a\v sik}
\author[2]{Jakub Cimerman}
\affil[1]{Univerzita Mateja Bela, Tajovsk\' eho 40, 97401 Bansk\' a Bystrica, Slovakia and \newline
FNSPE, \v Cesk\' e vysok\' e u\v cen\' i technick\' e v Praze, B\v rehov\'a 7, 11519 Praha 1, Czechia; boris.tomasik@cern.ch}
\affil[2]{FNSPE, \v Cesk\' e vysok\' e u\v cen\' i technick\' e v Praze, B\v rehov\'a 7, 11519 Praha 1, Czechia; jakub.cimerman@fjfi.cvut.cz}

\date{}

\begin{document}

\maketitle

\begin{abstract}
Event Shape Sorting is a novel method which is devised to organise a sample of collision events in such a way, that events with similar final state distribution of hadrons end up sorted close to each other. Such events are likely to have evolved similarly. Thus the method allows to focus at finer features of the collision evolution  because it would allow for averages over similar events that do not wash away these features. The algorithm is shortly explained. We also point out the distinction of Event Shape Sorting from the well established technique of Event Shape Engineering.

\end{abstract}


\section{Introduction}

Every fireball created in ultrarelativistic heavy-ion collisions is unique. The differences between fireballs are not merely statistical. Hadrons come from different statistical distributions in different events. This is firstly because of the impact parameter, which fluctuates from event to event, and secondly due to quantum fluctuations which are naturally present. The observed distribution of hadrons reflects the state of  the fireball at the moment of kinetic freeze-out. Its temperature, viscosity, and anisotropic expansion are the most important features which determine the hadronic spectra in transverse momentum, azimuthal angle, and rapidity. Different final state hadron distributions thus point to different evolution history of the fireball, since the final state results from the evolution which includes expansion due to pressure, and cooling. The usual experimental approach is to measure hadron distributions over a large number of collision events in order to improve statistics and minimise statistical errors. Even if the samples over which the summation happens are selected carefully within the same centrality class, they still contain events which may differ dramatically from each other in their final states and therefore also in their evolution. It seems desirable to perform such summation only over event samples which contain events as similar in their final distributions as possible. One can expect that such events would have evolved in a similar manner and it would make better sense to measure average quantities over such samples. It is the purpose of Event Shape Sorting (ESS) to create such event samples. 

A similar motivation has lead to the design of the Event Shape Engineering (ESE) technique \cite{ese}. We will point out the difference between the two techniques, here. To set up the stage, let us shortly review how ESE works. There, events are selected with respect to a sorting observable that must be defined in the particular analysis. To give a concrete example, this can be the size of the $\vec q_2$ vector, which measures the ellipticity of the azimuthal hadron distribution
\begin{equation}
q_2 = \left | \vec q_2 \right | = \frac{1}{\sqrt{M}} \left | \sum_{j=1}^{M} e^{2i\phi_j}\right |\,  ,
\end{equation}
where $M$ is the (sub)event multiplicity and $\phi_j$ the azimuthal angle of the $j$th particle. The engineered events are selected according to their values of $ q_2 $, usually one chooses some percentile of the totality of events with the largest or the smallest $q_2$. To avoid that the selection is only due to statistical fluctuation, the selection variable is measured on a subevent and the physics study is then performed on a different subevent. Let us stress that ESE can be done with respect to \emph{any} sorting observable, not just $ q_2 $.

In contrast to this, Event Shape Sorting is performed without the need to specify any sorting observable. The algorithm has been designed in \cite{andy} and its application to heavy-ion collisions was elaborated and described in \cite{renca}. We usually propose its application to the distribution of hadrons in azimuthal angle. It works iteratively and leads to such a sorting of events, that events with similar histograms in azimuthal angle end up close to each other. It is not specified beforehand, what property  of the distribution will turn out to be decisive for the sorting. The algorithm picks the dominant feature of the distribution by itself. On one hand, such a selection seems to be less controlled than in ESE, where the sorting variable must be pre-defined. On the other hand, ESS allows to \emph{discover} the dominant sorting features, even if they might be unexpected. 


\section{The Event Shape Sorting algorithm}

Let us explain how the algorithm works \cite{renca}. In general, events are sorted according to histograms in the selected observable. In our case this is most often the azimuthal angle, but it can be any other observable or even the histogram can be two or more-dimensional. Here we work out the example with the azimuthal angle.
\begin{enumerate}
\setcounter{enumi}{-1}
\item Since we usually do it in azimuthal angle, we first rotate all events so that their second-order event planes are aligned. Otherwise, two events may appear different in spite of being identical but rotated with respect to each other. In principle, this feature should be automatised in the algorithm, which would then somehow find the best match between the events, but this has not been fully developed and implemented, yet.
\item 
\label{s:jeden}
Sort all events according to the value of an observable which is thought to best characterise the shape of the studied histograms. This may be $q_2$, for example. In fact, this first sorting does not influence the final sorting of events, but if it is done smartly, it can speed up the iteration part of the algorithm that follows
\item 
\label{s:dva}
Divide the sorted sequence of events into percentiles of equal amounts of events. We'll use deciles, here. We refer to this percentiles as to event bins.
\item 
In each event bin, determine histograms summed over all events.
\item 
Now do a loop over all events in the sample. Every event can be characterised by its bin record $\{n_i\}$. Within the loop, do the following for every event: 
\begin{enumerate}
\item 
Determine the probabilities that it would be randomly drawn from a probability distribution given by summed histograms in each event bin. The calculation is based on Bayes' theorem and the formula is 
\begin{equation}
P(\mu|\{n_i\}) = \frac{\prod_i P(i|\mu)^{n_i} P(\mu)}{\sum_\nu \prod_i P(i|\nu)^{n_i} P(\nu)}\,  .
\end{equation}
Here, Greek symbols number the deciles and $i$ is the index for bins in azimuthal angle $\phi$. The priors $P(\mu) = 1/10$. The probability to find a particle in the $i$-th $\phi$-bin in an event from the $\mu$-th decile is 
\begin{equation}
P(i|\mu) = \frac{n_{\mu,i}}{M_\mu}\,  ,
\end{equation}
where $n_{\mu,i}$ is the number of particles in the $i$-th $\phi$-bin from all events in the $\mu$-th decile, and $M_\mu$ is the sum of all multiplicities from all events in decile $\mu$.
\item 
Determine the average
\begin{equation}
\bar\mu = \sum_\mu \mu \, P(\mu|\{ n_i\} )\,  .
\end{equation}
\end{enumerate}
\item 
Re-sort all events according to their values of $\bar \mu$.
\item 
If the order of events changed, return to step \ref{s:dva}. If not, the algorithm has converged. 
\end{enumerate}
It has been tested that irrespective of the initial sorting done in step \ref{s:jeden}, the final result is always the same. The only difference can be in reversing the order, but this must be considered as an identical result, since the algorithm is supposed to organise the events so that similar events end up close to each other. 







\section{What ESS can do and ESE cannot}

A question may be asked, if there is any result that one can achieve using ESS which cannot be obtained using ESE if the sorting variable is chosen appropriately. In other words: Is there any feature that Event Shape Sorting can find and Event Shape Engineering would fail to find? 

The idea that we want to present is to look at the correlation of the second-order  and the third-order event planes. Their correlation would influence the value of the correlator
\begin{equation}
\left \langle e^{i\{\theta_2 - \theta_3\}_{min}} \right \rangle = 
\int_0^\pi d\theta_2 \int_0^{2\pi/3}d\theta_3\,  \rho(\theta_2,\theta_3) \, e^{i\{\theta_2 - \theta_3\}_{min}}\, ,
\end{equation}
where $\langle \dots \rangle$ denotes event average, $\{ \dots \}_{min}$ is the smallest difference between event plane angles, and $\rho(\theta_2,\theta_3)$ is their distribution. For uncorrelated event planes,  $\rho(\theta_2,\theta_3)$ is just the normalisation constant and 
\begin{equation}
\left \langle e^{i\{\theta_2 - \theta_3\}_{min}} \right \rangle = \frac{3}{2\pi^2}\, 
\int_0^\pi d\theta_2 \int_0^{2\pi/3}d\theta_3\, e^{i\{\theta_2 - \theta_3\}_{min}} = \frac{3}{\pi}
\end{equation}

The trick is now, that such a value can also be obtained for \emph{correlated} event planes. Although somewhat artificial, one could think of a situation, where only two values for $(\theta_2 - \theta_3)$ are possible: $\delta$ and $-\delta$. We obtain
\begin{equation}
\left \langle e^{i\{\theta_2 - \theta_3\}_{min}} \right \rangle  = \frac{1}{2}  \left (  e^{i\delta}  + e^{-i\delta} \right ) = \frac{3}{\pi} \,  ,
\label{e:cr}
\end{equation}
which gives $\delta \approx 0.301$. If the event planes would be correlated so that they differ just by this angle, the correlator $\left \langle e^{i\{\theta_2 - \theta_3\}_{min}} \right \rangle$ would not reveal such a correlation and the situation would not be distinguished from the uncorrelated event planes. 

We simulated sets of 100,000 events with  the Blast-Wave Monte Carlo generator DRAGON \cite{dragon,dragon-u}, where we have set the temperature $T=120$~MeV, the transverse radius $R=7$~fm, the Bjorken longitudinal proper time at freeze-out $\tau = 10$~fm$/c$, and the transverse flow gradient $\rho_0 = 0.8$. Second-order anisotropies in shape and flow fluctuated within the intervals $a_2 \in (0,0.1)$ and $\rho_2 \in (0,0.1)$, respectively. Third-order anisotropies fluctuated in intervals $a_3 \in (0,0.03)$, $\rho_3 \in (0,0.03)$. Four different sets of events were generated in which the share of events correlated according to relation (\ref{e:cr}) was 100\%, 60\%, 30\%, and 10\%. Before ESS was performed on the events, they were rotated so that their second-order event planes were aligned. For comparison, ESE was also performed and $q_2$ was chosen as the selection variable. 

\begin{figure}[t]
\centering
\includegraphics[width=0.74\textwidth]{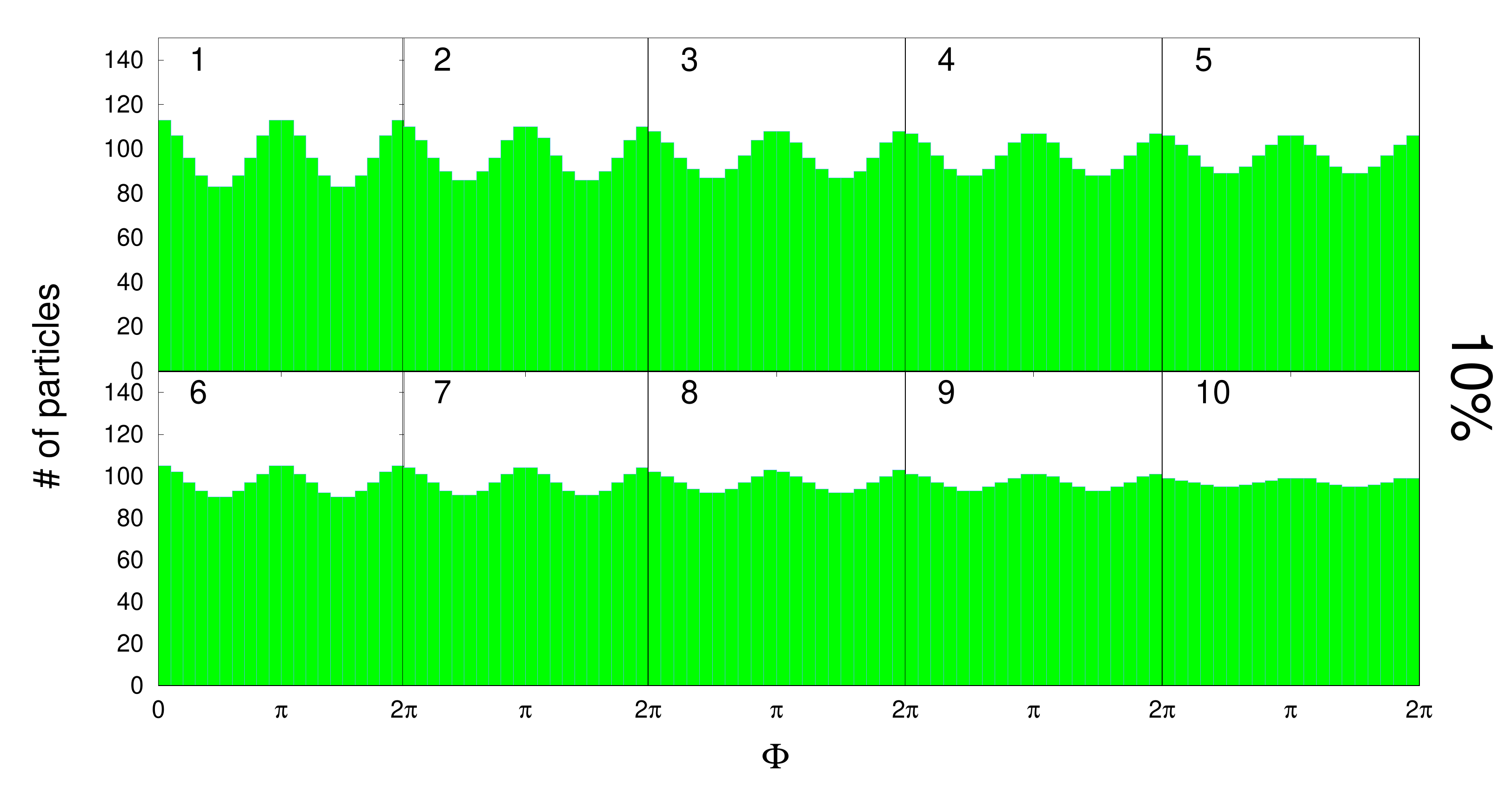}
\includegraphics[width=0.74\textwidth]{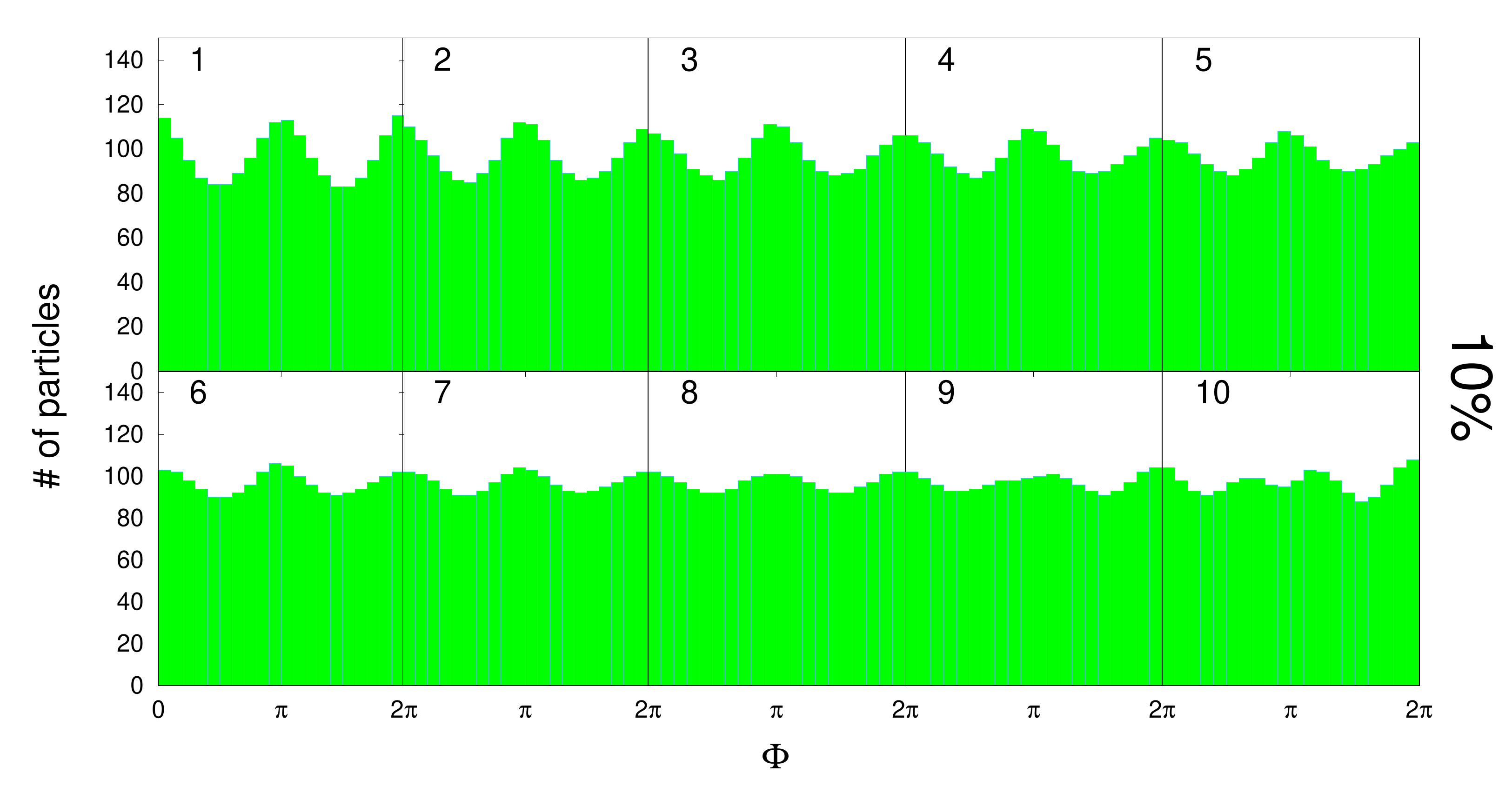}
\caption{Histograms of azimuthal hadron distributions from events divided into 10 event bins. Upper row: events are sorted and divided by ESE according to $q_2$ from large (event bin 1) to small (event bin 10). Lower row: events are sorted with ESS.}
\label{f:hist}
\end{figure}

As it was previously found in other studies, ESE did what it was expected to do: events were sorted according to the elliptic anisotropy, so that we can distinguish those with large $v_2$ from those with small $v_2$, see Fig.~\ref{f:hist}.
However, ESS organised the sequence of events in such a way that on one end of the sample events with clear elliptic anisotropy were collected, while on the other end of the sorted sample the shape is mainly determined by the third-order anisotropy. This kind of sorting may be hard to achieve with ESE, because the two different shapes are characterised by different dominant observables: the former by $q_2$ while the latter by $q_3$.  We should also note, however, that the histograms were constructed with event samples where only 10\% of all events have their second and third-order event planes correlated. We performed this analysis also for the other samples  where the partition of events with correlated event planes was larger. In those cases, the group of events with dominant third-order oscillation was not so pronounced, perhaps because it was more tightly coupled to the second-order oscillation. 

\begin{figure}[t]
\centering
\includegraphics[width=0.65\textwidth]{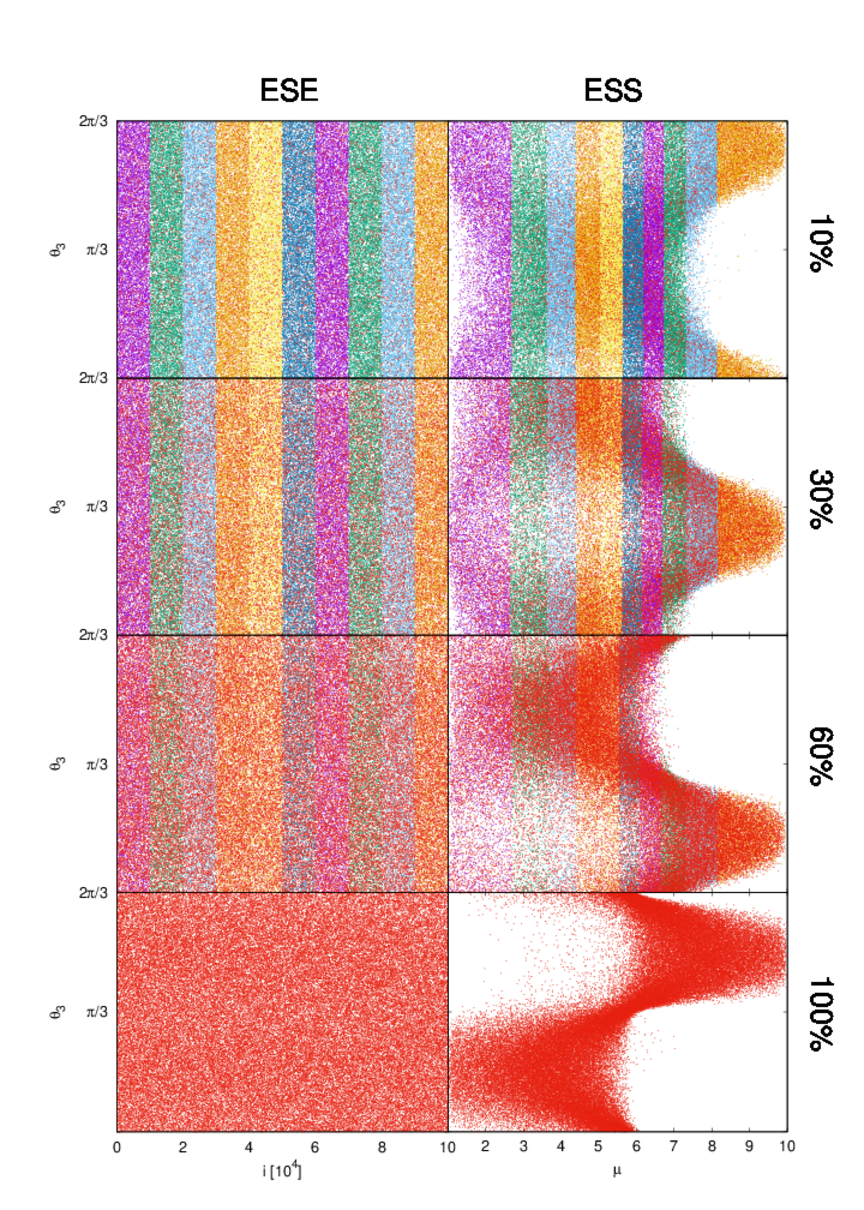}
\caption{Scatter plots show the relative angles $(\theta_2 - \theta_3)$ for each event as function of its sorting variable. Different rows correspond to different fractions of the events with correlated event planes, which are shown by black dots. Grey dots are event with uncorrelated event planes, while the change of the shade indicates a change of event bin. Left: Events are sorted by ESE with $q_2$ as the sorting variable. Right: Events are sorted with ESS. }
\label{f:scat}
\end{figure}

The important feature, however, hidden in our data, is the correlation between the event planes. Therefore, we check in Fig.~\ref{f:scat} how it is recognised by the algorithms. ESE is not sensitive to it, thus the selection of events based on the measurement of $q_2$ tells nothing about $\theta_2-\theta_3$. 
Recall that the correlation was constructed in such a way that in regular experimental analysis it would remain unobserved and one would not come to the idea to run ESE with this sorting variable. On the other hand, ESS recognises this feature very clearly. For the case with all events with correlation we see that they are divided basically into two groups with two different values of $\theta_2-\theta_3$. Such a feature still prevails if 60\% of all events contain the correlation. For smaller fractions it fades away, but there are always two to three event bins obtained by ESS which exhibit  the correlation between $\theta_2$ and $\theta_3$.  In fact, this may be naively expected, since 30\% of events would fill just three event bins and 10\% correspond to one event bin. Nevertheless, Fig.~\ref{f:scat} shows that the events with correlation are distributed among all  event bins, so that the division into event bins does not exactly distinguish event with and without correlation. 

In spite of that, we have found a feature which is identified by ESS while normally one would not recognise it in data analysis and thus ESE would not be set to look for it. 

\section{The influence of statistical fluctuations}

Statistical fluctuations are always present. A good sorting algorithm should not be sensitive to them. For this reason, in ESE events are usually selected according to a selection variable measured only on a subset of all particles observed in an event. The subsequent physics analysis is then performed in a different subset. Here, we want to test this approach on ESS. 

To this end we simulated two sets of events:
\begin{description}
\item [DRAGON] \cite{dragon,dragon-u} With this MC representation of the blast wave model we generated 150\,000 events with the same setting as in previous chapter.
\item [AMPT] \cite{ampt} With this transport model we have produced 10\,000 events of Au+Au collisions at the energy $\sqrt{s_{NN}} = 200$~GeV and with impact parameters between 7 and 10~fm. 
\end{description}
We have performed ESS on reference particles in each event and measured anisotropic flow on test particles, which do not include any reference particles. Our choice was
\begin{itemize}
\item reference particles: charged hadrons with rapidities $|y| < 0.4$;
\item test particles: charged hadrons with rapidities $0.5 < |y| < 0.8$. 
\end{itemize}

\begin{figure}[t]
\centering
\includegraphics[height=0.3\textheight]{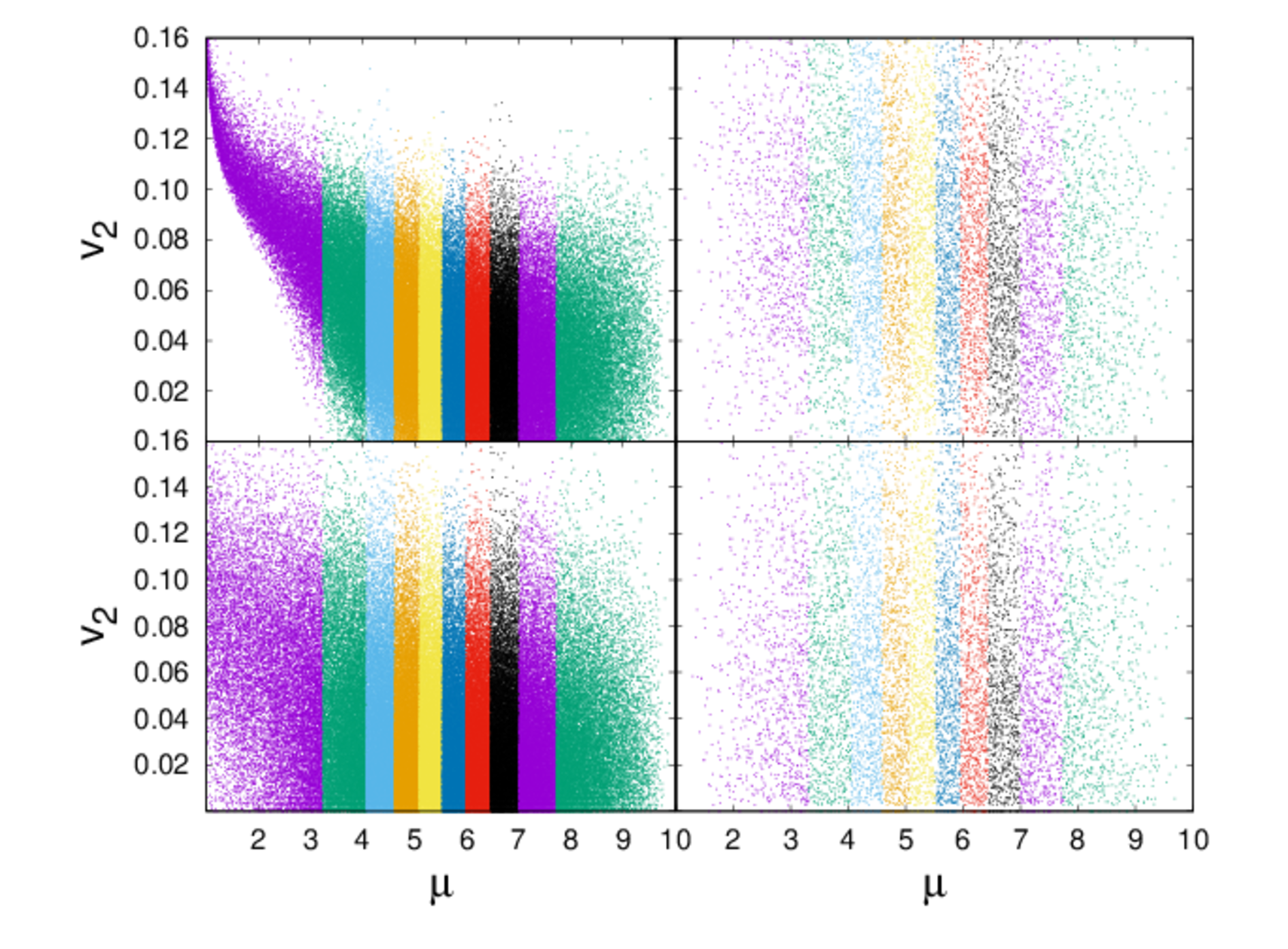}
\caption{Scatter plots show the distribution of events in variables $\bar\mu$ and $v_2$. Left column: events generated by DRAGON; right column: events generated by AMPT. Upper row: $v_2$ measured on reference particles;  lower row: $v_2$ measured on test particles.}
\label{f:stat}
\end{figure}

For comparison, we measured $v_2$ on test particles as well as reference particles. Results are shown in Fig.~\ref{f:stat}.
%
%
It is observed that for DRAGON events the sorting on reference particles was dominated by the elliptic anisotropy only for event bins with small $\mu$. This indicates that the shapes of the histograms have a richer structure, not just second-order anisotropy which was investigated here. Nevertheless, almost no signals of the $v_2$ dominance in the small-$\mu$ bins survive if it is measured on the test particles.  A null result on test particles would indicate that sorting on reference particles was only due to statistical fluctuations. However, we know how the events were generated and thus we know that there were non-statistical differences in second as well as third-order anisotropies between the events. In this figure they do not seem to show up. Currently, we are not sure about the reason why they remained hidden, so we draw no definite conclusions, yet. 

This standpoint is further supported by the analysis of AMPT events. In this case, we do not even observe clear $v_2$ distinction on the reference particles. Note that ESE would have sorted the events according to $v_2$ if it was asked to do so. No $v_2$ distinction is then naturally also expected for test particles. We speculate that small statistics may be among the reasons for this observation, although neither here we have a definite answer if and how to recognise statistical fluctuations. 

\section{Conclusions}

Event Shape Sorting can help to select events in such a way, that it creates samples of events which have similar momentum distribution of produced particles. According to what we know today about their evolution, such events likely evolved from similar initial conditions and followed similar expansion. 

One could argue that any measurable feature of the events can be identified and sorted out by the established technique of Event Shape Engineering. Nevertheless, we have found an example of a feature that would hardly be identified and used in ESE. Even though it might seem somewhat artificial, let us take it as an example that in principle something like this is possible. In real experiment, some other unexpected feature may be present---and unexpected features are hard to expect. ESS is versatile enough to catch them. 

The influence of statistical fluctuations on the method still needs to be better investigated.  

\paragraph*{Acknowledgment}
This work was supported by the grant 17-04505S of the Czech Science Foundation (GA\v CR) and by VEGA (Slovakia) under grant No. 1/0348/18.


\end{document}